\documentclass[prd,twocolumn,superscriptaddress,floatfix,amsmath,amssymb,amsfonts,nofootinbib,longbibliography]{revtex4-2}

\usepackage{float} 
\usepackage{scalerel}
\usepackage[normalem]{ulem}
\usepackage[english]{babel}
\usepackage{graphicx}
\usepackage{dcolumn}
\usepackage{bm}
\usepackage{blindtext}
\usepackage{verbatim}
\usepackage{relsize}
\usepackage{mathrsfs}
\usepackage{musicography}
\usepackage{amsmath}
\usepackage{blindtext}
\usepackage{cancel}
\usepackage{physics}
\usepackage{epstopdf}
\usepackage{mathtools}
\usepackage{blindtext}
\usepackage{tensor}
\usepackage{color}
\usepackage[usenames,dvipsnames]{pstricks}
\usepackage{epsfig}
\usepackage{pst-grad} 
\usepackage{pst-plot} 
\usepackage{hyperref}
\usepackage{verbatim}
\usepackage{slashed}
\usepackage{dsfont}
\usepackage[english]{babel}
\usepackage{amsmath,amssymb,amsthm}  

\newcommand{\mf}{\mathsf}

\newcommand{\ii}{\mathrm{i}}

\newcommand{\tc}[1]{\textsc{#1}}
\newtheorem{thm}{Theorem}

\allowdisplaybreaks[1] 

\begin{document}

\title{Duality between amplitude and derivative coupled particle detectors in the limit of large energy gaps}

\author{T. Rick Perche}
\email{trickperche@perimeterinstitute.ca}

\affiliation{Department of Applied Mathematics, University of Waterloo, Waterloo, Ontario, N2L 3G1, Canada}
\affiliation{Institute for Quantum Computing, University of Waterloo, Waterloo, Ontario, N2L 3G1, Canada}
\affiliation{Perimeter Institute for Theoretical Physics, Waterloo, Ontario, N2L 2Y5, Canada}

\author{Matheus H. Zambianco}
\email{mhzambia@uwaterloo.ca}

\affiliation{Department of Applied Mathematics, University of Waterloo, Waterloo, Ontario, N2L 3G1, Canada}
\affiliation{Institute for Quantum Computing, University of Waterloo, Waterloo, Ontario, N2L 3G1, Canada}

\begin{abstract}
    We present a duality between a particle detector model coupled to the amplitude of a scalar field and coupled to the field's derivative in the limit of large energy gaps. We show that the results of the models can be mapped to each other in a one-to-one fashion modulo a rescaling by the detector's gap. Our analysis is valid for arbitrary scalar fields in curved spacetimes, and requires minimal assumptions regarding the detectors. The duality also applies to the case where more than one detector is coupled to the field. This shows that many examples of entanglement harvesting with amplitude coupled UDW detectors give exactly the same result as derivative coupled detectors that interact with the field in the same region of spacetime.
\end{abstract}

\maketitle

\section{Introduction}

Dualities in theoretical physics uncover connections between seemingly disparate theories, bringing new perspectives and approaches to problems in different fields. Examples such as the holographic principle, with its AdS/CFT correspondence~\cite{Maldacena1997,VeronikaAds}, and T-duality in string theory~\cite{Tduality} demonstrate the profound impact of dualities on our comprehension of gravity and quantum mechanics. Moreover, dualities extend beyond high-energy physics, finding relevance in condensed matter through phenomena like topological insulators~\cite{topologicalDbranes,topologicalInsulators} and analogue gravity examples which can simulate quantum fields in curved spacetimes~\cite{SilkeTank,analBH,Superradiance}. These correspondences reveal a mathematical unity beneath the complexity of the physical world, and have the potential to offer novel approaches to long-standing puzzles.

While many dualities may connect areas of physics that are a priori vastly different, in this manuscript we will be concerned with a specific duality in the context of local interactions with quantum field theories. More precisely, dualities between two different kinds of couplings that localized systems might have with quantum fields---either by directly coupling to the field amplitude or by coupling with its derivative. Both of these types of couplings can be realized by physical systems coupled to quantum fields. For instance, atoms can be considered to either couple to the electromagnetic potential, or effectively couple to its derivative (the electric field) in different regimes~\cite{Nicho1}. Our results then establish a duality between two identical systems such that one locally interacts with a quantum field's amplitude and the other one with its derivative.

Systems that are locally coupled to QFTs have many different theoretical and experimental applications in fields that range from high-energy physics to quantum information. Overall, non-relativistic quantum systems coupled to quantum fields can be used to explore quantum information protocols in quantum field theory, even in curved spacetimes. These systems are usually termed particle detector models, and are the main tool of study of the field of relativistic quantum information (RQI). The applications of particle detectors in the study of fundamental aspects of quantum field theory are vast. Examples are studies of the Unruh effect~\cite{Unruh1976,Takagi,matsasUnruh,LoukoUnruh,antiUnruh,barbara,mine}, Hawking radiation~\cite{Unruh1976,JormaHawking,JormaHowThermal,aLittleExcit}, measurements in QFT~\cite{chicken,hectorChicken}, spacetime structure~\cite{geometry,ahmed,ericksonFinn}, thermodynamics in QFT~\cite{prlThermo,adam}, among others. Particle detectors have also inspired new quantum information protocols within quantum information theory, such as entanglement harvesting~\cite{Valentini1991,reznik2,Reznik2003,Pozas-Kerstjens:2015,Pozas2016,HarvestingSuperposed,HarvestingDelocalized,ericksonNew,ericksonBH,diki}, quantum collect calling~\cite{Jonsson2,collectCalling,PRLHyugens2015} and quantum energy teleportation~\cite{teleportation,teleportation2014}.

Due to their vast applications, the term ``particle detector'' is a very general term, which can be used to describe different systems, such as atoms interacting with the electromagnetic field~\cite{Pozas2016,Nicho1,richard}, neutrinos interacting with nucleons~\cite{neutrinos,antiparticles} and even the interaction of localized systems with linearized quantum gravity~\cite{pitelli,boris,ourBMV}. This leads to a variety of couplings with different quantum fields. Nevertheless, it has been shown that in specific cases, most of these couplings can be mimicked by simpler couplings of two-level systems with a scalar field~\cite{Pozas2016,richard,antiparticles}, and usually it is possible to generalize results from this case to more complex scenarios. However, even in the case of localized quantum systems coupled to a scalar field, there are different couplings that one can consider. For instance, a detector can be coupled to a scalar field $\hat{\phi}(\mf x)$ itself, or with its conjugate momentum $\hat{\pi}(\mf x)$, which will be the relevant cases in this manuscript.

We will study the relationship between particle detectors linearly coupled to the amplitude of quantum field, or coupled to its derivative. We will classify the regimes where these models give approximate results, and we will show that in the limit of large enough energy gaps for the detectors, the models are completely dual to each other under very reasonable assumptions. The situations where the models give approximate results match many cases considered in the literature, such as Gaussian switching functions and detectors with large enough gaps. Moreover, we will also address the duality in the case where more than one detector is coupled to the field. In the large gap limit, we will establish a duality between quantum information protocols implemented by amplitude coupled particle detectors and derivative coupled particle detectors.


Our manuscript is organized as follows. In Section \ref{sec:PD} we review particle detectors linearly coupled and coupled to the derivative of the field. In Section \ref{sec:duality} we establish the duality between the models, relating the results of the derivative coupled case with the amplitude coupled model, and we prove a theorem that shows that in the large gap limit, the two models behave similarly apart from a rescaling of the switching function. In Section \ref{sec:examples} we show that there are physically realizable energy gaps which allow the duality to approximately hold. We then conclude that many previous examples studied in the literature where the amplitude coupling was considered would give similar results for the derivative coupled model. In Section \ref{sec:harvesting} we review the protocol of entanglement harvesting, so that we can later compare the two scenarios. In Section \ref{sec:harvestingDuality} we establish this duality in the case where two detectors are coupled to the field, and discuss the implications in entanglement harvesting. The conclusions of our work can be found in Section \ref{sec:conclusions}.

\section{Particle Detector Models}\label{sec:PD}

\subsection{UDW model with field amplitude coupling}\label{sub:UDW}

In this section, we review the amplitude coupled two-level Unruh-DeWitt (UDW) model ~\cite{Unruh1976,DeWitt}, which is the simplest and most common particle detector model in the literature. This model has proven to be a powerful tool for implementing local operations in quantum fields, and allows one to easily consider quantum information protocols in quantum field theory. More recently, its applications have extended to probing spacetime structure~\cite{topology,geometry,ahmed}, addressing the problem of measurements in QFT~\cite{chicken}, and quantifying the entanglement present in quantum fields~\cite{kelly}. 

In its simplest form, the UDW model consists of a two-level system linearly coupled to a real scalar quantum field. In order to introduce the essential aspects of the model, let us consider an $n + 1$ dimensional globally hyperbolic spacetime ${\mf M}$ possessing a Lorentzian metric ${\mf g}$. The field $\hat{\phi}(\mf x)$ is assumed to satisfy the Klein-Gordon equation
\begin{equation}
    (\nabla^{\mu}\nabla_{\mu} - m^2  - \xi R) \hat{\phi}(\mf x) = 0,
    \label{KG_eq}
\end{equation}
where $m$ is the mass of the field, $\xi$ is a constant, and $R$ is the Ricci scalar. If $\{u_{\bm k}(\mf x), u_{\bm k}(\mf x)^*\}$ is a Klein-Gordon orthonormal basis of solutions for Eq. \eqref{KG_eq}, then we write
\begin{equation}
    \hat{\phi}(\mf x) = \int{\dd^{n}{\bm k} \  (u_{\bm  k}(\mf x)\hat{a}_{\bm k} + u_{\bm k}(\mf x)^{*}\hat{a}_{\bm k}^{\dagger})},
\end{equation}
where the ladder operators satisfy the bosonic canonical commutation relations, namely
\begin{equation}
    [\hat{a}_{\bm k}, \hat{a}_{\bm p}^{\dagger}] = \delta^{(n)}(\bm k - \bm p).
\end{equation}

As for the detector, we consider it to be following a spacetime trajectory $\mf z(\tau)$ in ${\mf M}$, with $\tau$ being the proper time of the trajectory. Furthermore, we denote by $\{\Sigma_{\tau}\}_{\tau \in \mathbb{R}}$ the local rest spaces of the trajectory $\mf z(\tau)$ that can be covered by Fermi normal coordinates (FNC) ${\mf x}=(\tau, {\bm x})$ and locally foliate ${\mf M}$~\cite{poisson,us,us2}.

The quantum degrees of freedom of the detector are described as a two-level system whose time evolution is defined by the following Hamiltonian:
\begin{equation}
    \hat{H}_{\tc{d}} = \Omega \hat{\sigma}^{+} \hat{\sigma}^{-}.
\end{equation}
Here, $\Omega$ is the energy gap, whereas $\hat{\sigma}^{+}$ and $\hat{\sigma}^{-}$ are the usual ladder operators. For the sake of concreteness, if the Hilbert space of the system is spanned by the orthonormal states $\{|g \rangle, |e \rangle\}$ (the ground and excited states), then we write $\hat{\sigma}^{+} = |e \rangle \langle g|$ and $\hat{\sigma}^{-} = |g \rangle \langle e|$.

The coupling between the field and the detector is 
completely characterized by the interaction Hamiltonian density (written in the interaction picture)
\begin{equation}
    \hat{h}_{\tc{I}}(\mf x) = \lambda \Lambda(\mf x) \hat{\mu}(\tau)\hat{\phi}(\mf x).
    \label{h_I}
\end{equation}
In the equation above, $\lambda$ is a coupling constant, and $\Lambda(\mf x)$ is the so called spacetime smearing function, which allows us to consider a more realistic model where the interaction between the field and the detector is localized in a finite region of spacetime. Indeed, it is common to split this function as $\Lambda(\mf x) = \chi(\tau) f(\bm x)$, where $\chi(\tau)$ is the switching function, which controls the process of switching the detector's interaction ``on'' and ``off'', and $f(\bm x)$ describes the shape of the detector in its rest frame. The fact that $f$ does not depend on $\tau$ imposes the rigidity condition for the detector shape~\cite{us,mine}. The operator $\hat{\mu}(\tau)$ is called the detector's monopole operator, and in the interaction picture it is given by
\begin{equation}
    \hat{\mu}(\tau) = e^{\ii \Omega \tau} \hat{\sigma}^{+} + e^{-\ii \Omega \tau} \hat{\sigma}^{-}.
    \label{mu_tau}
\end{equation}
It will be relevant for us to study the dimensions of $\lambda$, $\Lambda(\mf x)$ and $\hat{\phi}(\mf x)$. With conventions such that $\hbar = c = 1$, we impose that $\Lambda(\mf x)$ has the units of a spatial density, which corresponds to units of 
$E^n$, where $E$ is a unit of energy. We then find that $[\hat{\phi}] = E^{\frac{n-1}{2}}$, which gives us $[\lambda] = E^{\frac{3-n}{2}}$. This particular choice makes it so that the coupling constant is dimensionless only in $n=3$ spatial dimensions.

The evolution of states in the total system (field + detector) is governed by the time evolution operator
\begin{equation}
    \hat{U}_{I}(\tau_{1}, \tau_{0}) = {\cal T}_{\tau} \exp \left( -\ii\int_{\mf M_{\tau_0 \tau_1}}{\!\!\!\!\!\!\!\!\dd V  \hat{h}_{\tc{I}}(\mf x)}\right),
    \label{U_I}
\end{equation}
where $\mf M_{\tau_0\tau_1}$ is the spacetime region between the surfaces ${\Sigma_{\tau_0}}$ and ${\Sigma_{\tau_1}}$, and $\dd V = \dd \tau \dd^{n}\bm x \sqrt{-g}$ is the invariant spacetime volume element. A word of attention is in order here: as shown in~\cite{us2}, in general, the time ordering operator will depend upon the choice of time parameter and foliation between the surfaces $\Sigma_{\tau_0}$ and $\Sigma_{\tau_1}$. However, provided we keep all our predictions to leading order in the perturbative parameter $\lambda$ and choose suitable initial states, there will be no dependence in the notion of time ordering chosen. Thus, we can basically write ${\cal T}$ instead of ${\cal T}_{\tau}$ in this context, and write the time ordering operation with respect to any parameter to leading order in perturbation theory.

Our main interest is to compute the final state of the probe after the interaction with the field. In order to do so, let us consider the initial state
\begin{equation}
    \hat{\rho}_{0} = |g \rangle \langle g| \otimes \hat{\rho}_{\phi, 0},
    \label{rho_0}
\end{equation}
where we assume $\hat{\rho}_{\phi, 0}$ to be any Gaussian state of the quantum field.

The state $\hat{\rho}_{0}$ then evolves with $\hat{U}_{\tc{I}} = \hat{U}_{\tc{I}}(\infty, -\infty)$. We take $\tau_0\to - \infty$ and  $\tau_{1} \to \infty$ because the spacetime support of the spacetime smearing function $\Lambda(\mf x)$ automatically implements the finite time duration of the interaction\footnote{Typically, when one writes $\Lambda(\mf x) = \chi(\tau)f(\bm x)$, one assumes $\chi(\tau)$ and $f(\bm x)$ to be localized functions which define a finite (or approximately finite) interaction region.}. 
 
After tracing out the field degrees of freedom, we have the density matrix  $\hat{\rho}_{\tc{d}}$ that describes the final state of the detector. To leading order in $\lambda$, we find
\begin{equation}
    \hat{\rho}_{\tc{d}} = \Tr_{\phi}[\hat{U}_{I} \hat{\rho}_{0} \hat{U}_{I}^{\dagger}] = \begin{bmatrix}
{ 1 - \cal L} & 0 \\
0 & {\cal L}  \\ 
\end{bmatrix}
+  {\cal O}(\lambda ^{4}),
\label{rho_D}
\end{equation}
with
\begin{equation}
    {\cal L} = \lambda^{2} \int{\dd V \dd V' e^{-\ii \Omega(\tau - \tau')} \Lambda(\mf x) \Lambda(\mf x') W(\mf x, \mf x')}.
\end{equation}
Notice that the next order would be $\lambda^4$ and not $\lambda^3$, due to the assumption that $\hat{\rho}_{\phi,0}$ is a Gaussian state. It can then be shown that all the predictions of the theory are functions of integrals of the  Wightman function
\begin{equation}
    W(\mf x, \mf x')_{\rho_{\phi, 0}} = \langle\hat{\phi}(\mf x)  \hat{\phi}(\mf x') \rangle_{\hat{\rho}_{\phi,0}} = \Tr[\hat{\phi}(\mf x)  \hat{\phi}(\mf x') \hat{\rho}_{\phi,0}].
    \label{W}
\end{equation}
When the context makes it clear over which state we are evaluating the correlation function, we will just write $W(\mf x, \mf x')$ for short.


\subsection{UDW model with derivative coupling}\label{sub:derUDW}

Many other types of coupling between the field and the detector have been explored in the literature. One of the most used models is the derivative coupling, where the detector couples to the derivative of the field. This model has been considered in the literature in many studies in both flat and curved spacetimes~\cite{ericksonBH,derivativeJorma,ElModeEstaLouko,bunnyCircular}. In fact, the derivative coupling in lower dimensions can be used to obtain particle detector models that mimic the response of linearly coupled detectors in higher dimensions, as considered in~\cite{ericksonBH,derivativeJorma,bunnyCircular}, for instance. The derivative coupling can also be used to mimic couplings of systems with the electromagnetic field, as discussed in~\cite{Pozas2016,richard}. The main goal of this manuscript is to establish a duality between the derivative coupling and the amplitude coupling models in the same spacetime of dimension $n+1$.

 In order to define the derivative coupled UDW model, consider $\partial_\tau$ to be the timelike coordinate vector field in FNC, defined locally around the trajectory of the detector $\mf z(\tau)$. Then, the interaction Hamiltonian density that defines the UDW model with derivative coupling is
\begin{equation}
    \hat{\Tilde{h}}_{\tc{I}}(\mf x) = \Tilde{\lambda} \Tilde{\Lambda}(\mf x) \hat{\Tilde{\mu}}(\tau)\partial_\tau\hat{\phi}(\mf x).
\end{equation}
Notice that we have used tildes to denote the operators and functions in this model. This notation will come in handy when we discuss the duality which is the main result of this paper. Furthermore, it is important to notice that the addition of the derivative to the field adds an extra dimension of energy that has to be compensated by one of the terms present in $\hat{\Tilde{h}}_I(\mf x)$, therefore changing their units. We will keep $\Tilde{\lambda}$ dimensionless, and change the units of the spacetime smearing function. That is, $[\Tilde{\Lambda}] = E^{n-1}$. In particular, in the case of $n = 3$, we find that $\Tilde{\Lambda}$ now has units of a two-dimensional density ($E^2$)\footnote{It is common in the literature to instead change the dimensions of the coupling constant in the case of the derivative coupling~\cite{ericksonBH,derivativeJorma,ElModeEstaLouko,bunnyCircular}. However, in this manuscript it will be more convenient to rescale the switching function $\Tilde{\chi}(\tau)$ instead, under the assumption that $\Tilde{\Lambda}(\mf x) = \Tilde{\chi}(\tau)\Tilde{f}(\bm x)$. It is important to note that regardless of the choice made, $\Tilde{\lambda} \Tilde{\chi}(\tau)$ will have the same units in both cases.}.

The final state for a UDW detector with derivative coupling can be computed by following exactly the same procedure as in the linear coupling case, with the replacement $\hat{\phi}(\mf x) \mapsto \hat{\psi}(\mf x) = \partial_\tau \hat{\phi}(\mf x)$. Indeed, if we consider the initial state of the detector-field system to be $\ket{g}\!\!\bra{g}\otimes \hat{\rho}_{{\phi},0}$, we can still write the final state of the detector as in Eq. \eqref{rho_D}, with the replacement $\mathcal{L}\mapsto\tilde{\mathcal{L}}$, which is defined by

\begin{equation}
    \Tilde{{\cal L}} =  \Tilde{\lambda}^{2} \int{\dd V \dd V' e^{-\ii \Omega(\tau - \tau')} \Tilde{\Lambda}(\mf x) \Tilde{\Lambda}(\mf x') \Tilde{W}(\mf x, \mf x')},
\end{equation}
where the function $\Tilde{W}(\mf x, \mf x')$ is the correlation function of the operator $\hat{\psi}(\mf x) = \partial_\tau \hat{\phi}(\mf x)$ in the state $\hat{\rho}_{\phi,0}$, namely
\begin{equation}
    \Tilde{W}(\mf x, \mf x') =  \langle\partial_{\tau}\hat{\phi}(\mf x) \partial_{\tau'}\hat{\phi}(\mf x') \rangle_{\hat{\rho}_{\phi, 0}} = \langle\hat{\psi}(\mf x) \hat{\psi}(\mf x') \rangle_{\hat{\rho}_{\phi, 0}}.
    \label{W_tilda}
\end{equation}
In essence, all calculations that can be done for the case of the amplitude coupling model can be mapped into the derivative coupling model by replacing $\hat{\phi}(\mf x) \mapsto \hat{\psi}(\mf x)$, and changing the corresponding spacetime smearing function and coupling constant. Notice that due to the different correlation functions, there is no a priori reason to expect the results to of each model to behave similarly.

\section{Duality between linear coupling and derivative coupling}\label{sec:duality}

In this section, we present the duality between the UDW model with amplitude coupling and the UDW model with derivative coupling. Such a duality will be characterized as a relationship between the spacetime smearing functions of both cases, with some special attention given to the switching function that controls the time window of the interaction between the field and the local probes. For the sake of notation clarity, quantities related to the derivative coupling case will be distinguished from the ones in the amplitude coupling case by the symbol $\sim$, as in Section \ref{sec:PD}. Nonetheless, in order to address a more general result, we will allow for the energy gap $\tilde{\Omega}$ to be time dependent. The effect of this will then be to change the free Hamiltonian for the derivative coupling to
\begin{equation}
    \hat{\Tilde{H}}_{\tc{d}} = \Tilde{\Omega}(\tau) \hat{\sigma}^{+} \hat{\sigma}^{-}.
    \label{H_D_tilde_time_dependent_gap}
\end{equation}
With this change, the interaction Hamiltonian density of the derivative coupling UDW detector (in the interaction picture) can be written as

\begin{equation}
         \hat{\Tilde{h}}_{\tc{I}}(\tau) =\tilde{\lambda}\Tilde{\Lambda}(\mf x) (  e^{\ii \theta(\tau)}\hat{\sigma}^{+} + e^{-\ii \theta(\tau)}\hat{\sigma}^{-})\partial_{\tau}\hat{\phi}(\mf x),
     \label{h_I_duality_tilda}
\end{equation}
where the Heisenberg equation gives the relationship between $\tilde{\Omega}(t)$ and $\theta(t)$:
\begin{equation}
    \dv{}{\tau}\theta(\tau) = \tilde{\Omega}(\tau).
    \label{d_theta}
\end{equation}

We now consider the detector to start in the ground state as in Eq. \eqref{rho_0} and evolve it through the time evolution operator \eqref{U_I} with the appropriate interaction Hamiltonian density in each case --- either $\hat{h}_{\tc{I}}(\tau)$ or $\hat{\Tilde{h}}_{\tc{I}}(\tau)$. After tracing out the field degrees of freedom, the density matrix describing the detector's state will have exactly the same form as Eq. \eqref{rho_D} for both the amplitude and derivative coupling cases. The only difference is that now, for the later case, we need to make the replacement $\mathcal{L}\mapsto\tilde{\mathcal{L}}$, with:

\begin{equation}
    \Tilde{{\mathcal{L}}} = \Tilde{\lambda}^{2} \int{\dd V \dd V'  \Tilde{\Lambda}(\mf x) \Tilde{\Lambda}(\mf x ')e^{- \ii (\theta(\tau) - \theta(\tau'))} \Tilde{W}(\mf x, \mf x')},
    \label{Tilda_L_duality}
\end{equation}
where $W(\mf x, \mf x')$ and $\Tilde{W}(\mf x, \mf x')$ are given by Eqs. \eqref{W} and \eqref{W_tilda}, respectively. In particular, they satisfy 
\begin{equation}\label{eq:key}
    \Tilde{W}(\mf x, \mf x') = \partial_\tau \partial_{\tau'}W(\mf x, \mf x').
\end{equation}

In order to establish a relationship between the final state in the two models, it is then enough to relate $\mathcal{L}$ and $\tilde{\mathcal{L}}$. In order to do so, we will perform integration by parts in the integral of Eq. \eqref{Tilda_L_duality} using the relationship between the correlation functions in Eq. \eqref{eq:key}. For convenience, let us write $\partial_\tau = u^\mu \nabla_\mu$ in what follows, so that we can apply the divergence theorem straightfowardly with the volume element $\dd V$.
\begin{align}
     \Tilde{{\mathcal{L}}} &= \Tilde{\lambda}^{2} \!\!\int{\dd V \dd V'  \Tilde{\Lambda}(\mf x) \Tilde{\Lambda}(\mf x ')e^{- \ii (\theta - \theta')} u^\mu u^{\nu'} \nabla_\mu \nabla_{\nu'}W(\mf x, \mf x')}\nonumber\\
     &= -\Tilde{\lambda}^{2}\!\! \int{\dd V \dd V'  \nabla_\mu(\Tilde{\Lambda}(\mf x)e^{-\ii \theta}u^\mu) \Tilde{\Lambda}(\mf x ')e^{\ii  \theta'} \! u^{\nu'}\nabla_{\nu'}W(\mf x, \mf x')}\nonumber\\
     &= \Tilde{\lambda}^{2} \!\!\int{\dd V \dd V'  \nabla_\mu(\Tilde{\Lambda}(\mf x)e^{-\ii \theta}u^\mu) \nabla_{\nu'}(\Tilde{\Lambda}(\mf x ')e^{\ii  \theta'} \! u^{\nu'}\!) W(\mf x, \mf x')},  \label{IBP}  
\end{align}
where we have neglected the boundary terms under the assumption that the spacetime smearing functions are localized.
In order to study the conditions under which the amplitude coupled model and the derivative coupled model will give the same result, we can now impose \mbox{$\mathcal{L} = \tilde{\mathcal{L}}$}, which will lead us to the following differential equation for $\tilde{\Lambda}(\mf x)$ and $\theta(\tau)$ in terms of $\Lambda(\mf x)$ and $\Omega$:
\begin{equation}
     \tilde{\lambda}\left(\partial_\tau(\Tilde{\Lambda}(\mf x)e^{-\ii \theta(\tau)}) +(\nabla_\mu u^\mu)\Tilde{\Lambda}(\mf x) e^{- \ii \theta(\tau)}\right) = \lambda\Lambda(\mf x) e^{- \ii \Omega \tau}.
     \label{duality_raw}
\end{equation}
If the equation above is satisfied for the spacetime smearing function $\Tilde{\Lambda}(\mf x)$ and the $\theta(\tau)$, the amplitude coupled model and the derivative coupled model will yield the same results\footnote{Importantly, $\tilde{\Lambda}(\mf x)$ has to be a localized function in order for the integration by parts to be performed.}.

Although Eq. \eqref{duality_raw} might look complicated at first, a few reasonable assumptions can be used in order to simplify it. First, we will assume that the vector field $\partial_\tau$ is divergenceless. That is, $\nabla_\mu u^\mu = 0$. This assumption naturally gets rid of one of the terms we have. Next, we assume the rigidity condition for both the amplitude and derivative coupled cases, so that $\Lambda(\mf x) = \chi(\tau)f(\bm x)$ and $\tilde{\Lambda}(\mf x) = \tilde{\chi}(\tau) \tilde{f}(\bm x)$. This allows us to factor out the space dependence in the equation. Under the assumption that $\tilde{\lambda} = \lambda$ and $\tilde{f}(\bm x) = f(\bm x)$ (that is, the detectors have the same coupling constant and the same rigid shape in the rest frame of $\mf z(\tau)$), Eq. \eqref{duality_raw} can be recast as
\begin{equation}
    \dv{}{\tau}\left(\tilde{\chi}(\tau)e^{-\ii \theta(\tau)}\right) = \chi(\tau) e^{-\ii \Omega \tau}.
\end{equation}
This equation can be directly integrated, and yields\footnote{Notice that while we can choose any bottom limit for the integral, the choice of $-\infty$ is convenient, as it ensures that $\lim_{\tau\to -\infty}\tilde{\chi}(\tau) = 0$.}
\begin{equation}
    \tilde{\chi}(\tau)e^{-\ii \theta(\tau)} = \int_{-\infty}^\tau\dd \xi  \chi(\xi) e^{-\ii \Omega \xi}.
    \label{chi_tilda_exp_theta}
\end{equation}
 In particular, under the assumption that $\tilde{\chi}(\tau)$ is a positive function, we find that $\theta(\tau)$ is the complex phase of the integral on the right-hand side of Eq. \eqref{chi_tilda_exp_theta}, while
\begin{equation}
    \tilde{\chi}(\tau) = \left|\int_{-\infty}^\tau\dd \xi  \chi(\xi) e^{-\ii \Omega \xi}\right|.
\end{equation}

Although we were able to relate the parameters of the derivative coupled theory to the parameters of the amplitude coupled model, we still have to check that there are cases where $\tilde{\chi}(\tau)$ is a function strongly supported in a finite region, so that the integration by parts of Eq. \eqref{Tilda_L_duality} can be performed with no boundary terms. In order to address this, we state and prove the following theorem.

\begin{thm}
    Let $\chi(t)\in L^1(\mathbb{R})$ be a differentiable function such that $\chi'(t)\in L^1(\mathbb{R})$ and $\lim_{t\to-\infty}\chi(t) = 0$.
Define
\begin{equation}
        f_{\Omega}(t) = \int_{-\infty}^t \dd \xi\,\chi(\xi) e^{- \ii \Omega \xi}.
        \label{f(t)_theo1}
\end{equation}
Then, the following relation holds:
\begin{equation}
    \lim_{\Omega\to \infty} \Omega f_{\Omega}(t) - \ii \chi(t) e^{-\ii \Omega t} = 0.
    \label{theorem1}
\end{equation}
\end{thm}
\vspace{1mm}
\begin{proof}
Since $\chi(t) \in L^1(\mathbb{R})$ is a differentiable function, we can perform integration by parts in Eq. \eqref{f(t)_theo1}. We obtain

\begin{align}
    \Omega f_{\Omega}(t) - \ii\chi(t)e^{-\ii\Omega t} &= -\ii\int_{-\infty}^{t}{d \xi \ e^{-\ii \Omega \xi} \chi'(\xi)} \label{theorem1_eq1}\\
    &= -\ii\int_{-\infty}^{\infty}{d \xi \ e^{-\ii \Omega \xi} \chi'(\xi)\theta(\xi - t)} .\nonumber    
\end{align}
Now notice that for each fixed $t\in\mathbb{R}$, we can apply the Riemann-Lebesgue lemma to the right-hand side of Eq. \eqref{theorem1_eq1}, which gives us the desired result.
\end{proof}

Now, using Theorem 1 in Eq. \eqref{chi_tilda_exp_theta}, we find that for large values of the energy gap $\Omega$,
\begin{equation}
    \tilde{\chi}(\tau)e^{-\ii \theta(\tau)} \approx \frac{\ii}{\Omega}\chi(\tau) e^{\ii \Omega \tau}.
\end{equation}
The equation above gives us the relationship between $\tilde{\chi}(\tau)$ and $\chi(\tau)$ and the relationship between $\theta(\tau)$ and $\Omega$:
\begin{equation}
    \Tilde{\chi}(\tau) \approx \frac{\chi(\tau)}{\Omega}, \quad \theta(\tau) = \Omega \tau + \frac{\pi}{2} \Rightarrow \tilde{\Omega}(\tau) = \Omega.\label{dualityValues}
\end{equation}
That is, we find that apart from a rescaling due to the energy gap, we have the exact same behaviour for $\tilde{\chi}(\tau)$ and $\chi(\tau)$, so that Eq. \eqref{IBP} holds in this case. We also find that in the limit of large energy gaps, the possibly time dependent energy gap $\tilde{\Omega}$ is independent of $\tau$, and matches the constant gap of the amplitude coupled model, $\Omega$. Also notice that this matches our conventions for units in the spacetime smearing function stated in Subsection \ref{sub:derUDW}, where $\Tilde{\Lambda}(\mf x)$ has one less dimension of energy than $\Lambda(\mf x)$. 

The conclusion is that in the limit of large gaps, the following Hamiltonian density for a derivative coupled theory
\begin{equation}
    \hat{\tilde{h}}_I(\mf x) = \frac{1}{\Omega}\lambda \chi(\tau)(e^{\ii \Omega \tau}\hat{\sigma}^+ + e^{-\ii \Omega \tau}\hat{\sigma}^-)f(\bm x)\partial_\tau\hat{\phi}(\mf x)
\end{equation}
will produce exactly the same results as one would obtain using the interaction Hamiltonian of Eq. \eqref{h_I}. Intuitively, this means that for sufficiently large gaps, the effective field probed by the detector is such that $\partial_\tau\hat{\phi}(\mf x)\approx \Omega\hat{\phi}(\mf x)$, which is to say that the frequency of the field which would be the most relevant for the detector is the frequency which resonates with its energy gap $\Omega$.

Overall, provided that the following three conditions hold:
\begin{enumerate}
    \item Rigidity: $\Lambda(\mf x) = \chi(\tau)f(\bm x)$ and $\tilde{\Lambda}(\mf x) = \tilde{\chi}(\tau) \tilde{f}(\bm x)$.\label{cond1}
    \item Constant local spacetime volume: $\nabla_\mu u^\mu = 0$.\label{cond2}
    \item Large energy gaps: $\Omega\to \infty$,\label{cond3}
\end{enumerate}
we have shown that the amplitude and derivative coupled UDW detectors will give the exact same results, with the rescaling of the interaction: $\tilde{\lambda} \Omega\tilde{\Lambda}(\mf x) = \lambda \Lambda(\mf x)$.

The result of this section implies that there are localized and physically realizable switching functions for the derivative coupling case which are dual to the amplitude coupled theory.
However, the exact condition that we found so that these switchings can be defined is $\Omega\to \infty$. A quantum system with an infinite energy gap is unphysical, so the natural question to be asked is whether there are finite values of $\Omega$ that will allow the condition $\Omega \tilde{\chi}(\tau) \approx \chi(\tau)$ to hold. It turns out that there are relevant examples where this is the case, and we will discuss these in Section \ref{sec:examples}. In particular, we will see that for a Gaussian switching function with standard deviation $T$, values of $\Omega$ as small as $6/T$ are enough to make the duality approximately hold.

Notice that the duality presented in this section holds for an arbitrary scalar quantum field in an arbitrary curved spacetime of dimension $n+1$. This includes the case where the field's correlation function might require an IR regulator. After the regulator is prescribed, and $W(\mf x, \mf x')$ is well defined, the duality presented in this section will be valid for sufficiently\footnote{In this case the spacetime smearing functions have to decay faster than the growth of the Wightman function with the spacetime separation between the events. This is in order to ensure that the excitation probability integrals are convergent.} localized probes.

Finally, notice that although we have carried on our analysis with a specific choice of initial state for the detector, it could have been carried on with any other initial detector state, and the duality would also hold. Indeed, the key point is that it is possible to integrate by parts terms which are integrals of correlation functions of the field. Given that all predictions from particle detector models can be written in terms of these, the duality between amplitude and derivative coupling holds for any choice of initial state of the detector.

\section{Examples}\label{sec:examples}

In this section we will discuss specific examples of switching functions and analyze whether there are physically realistic values of the energy gap that give localized switching functions $\Tilde{\chi}(\tau)$ for finite values of $\Omega$. In particular, we will analyze the result of Theorem 1, and check how large $\Omega$ has to be in comparison to the other scales so that $\Omega\Tilde{\chi}(\tau) \approx \chi(\tau)$ and $\Tilde{\Omega}(\tau)\approx \Omega$.

\begin{figure}[h!]
    \centering
    \includegraphics[width=8.6cm]{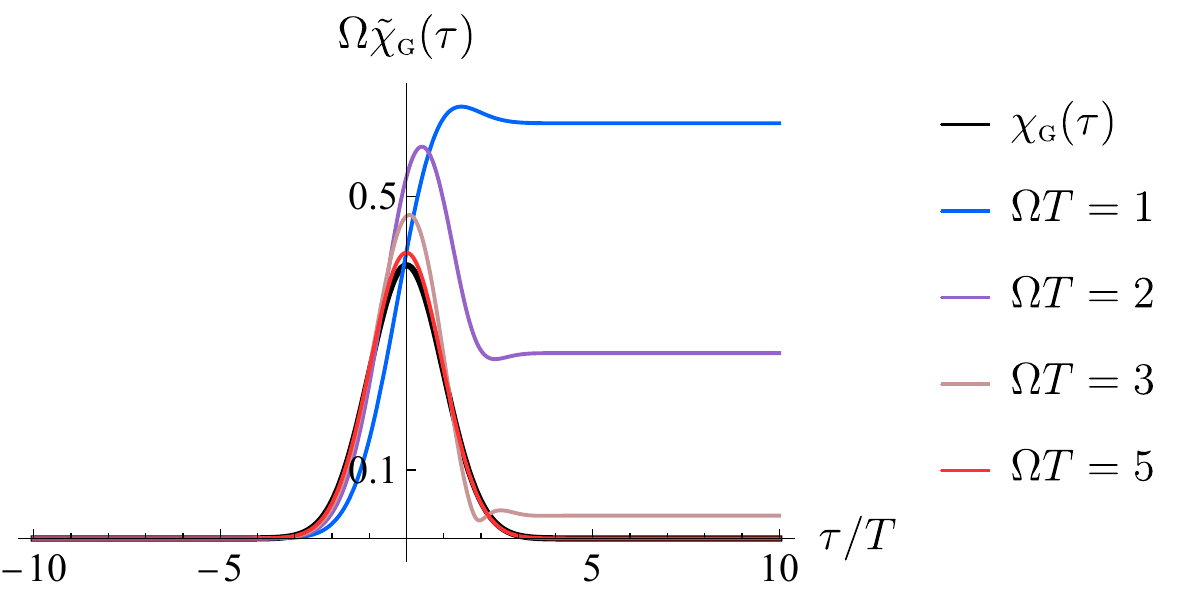}
    \caption{Analysis of the convergence of $\Omega \Tilde{\chi}_\tc{g}(\tau)$ to $\chi_\tc{g}(\tau)$ with the choice of Gaussian switching function of Eq. \eqref{chiG}.}
    \label{fig:Gaussian}
\end{figure}

\begin{figure}[h!]
    \centering
    \includegraphics[width=8.6cm]{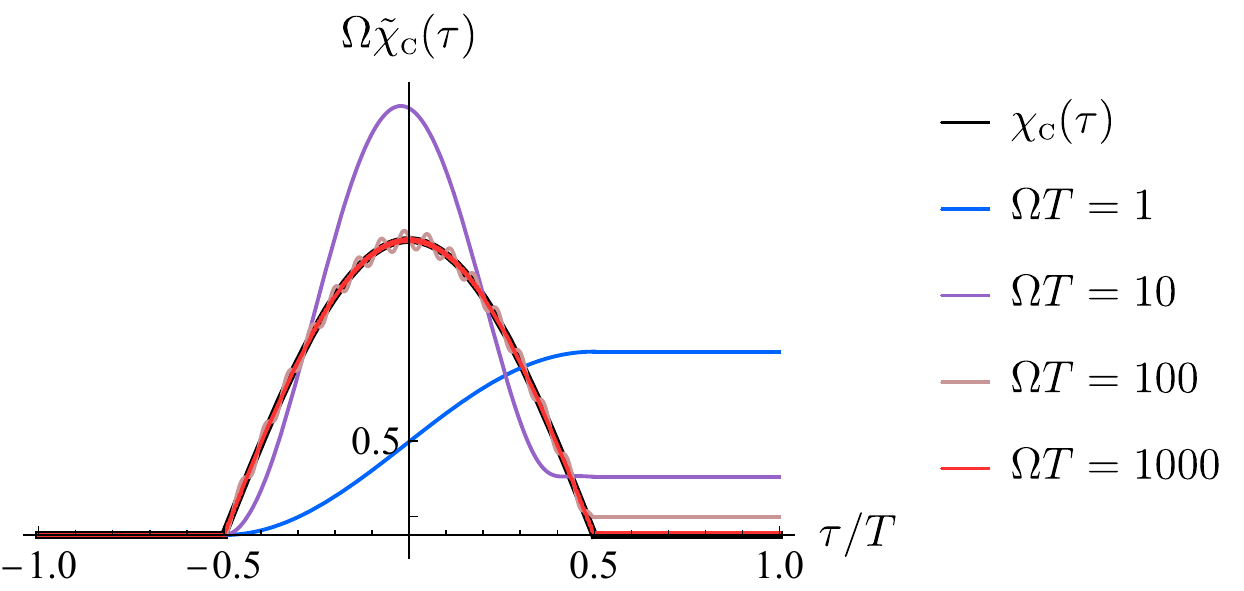}
    \caption{Analysis of the convergence of $\Omega \Tilde{\chi}_\tc{c}(\tau)$ to $\chi_\tc{c}(\tau)$ with the choice of compact switching function of Eq. \eqref{chiC}.}
    \label{fig:Compact}
\end{figure}

\begin{figure}[h!]
    \centering
    \includegraphics[width=8.6cm]{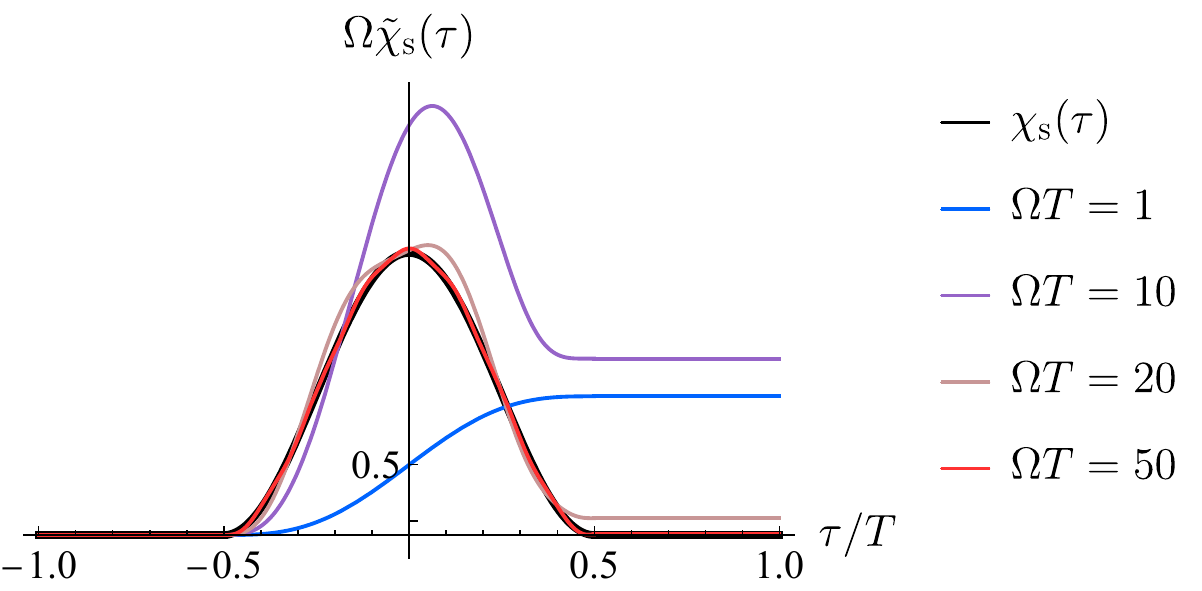}
    \caption{Analysis of the convergence of $\Omega \Tilde{\chi}_\tc{s}(\tau)$ to $\chi_\tc{s}(\tau)$ with the choice of compact switching function with continuous derivatives of Eq. \eqref{chiS}.}
    \label{fig:CompactSmooth}
\end{figure}

We will focus in three typical examples considered in the literature for the study of finite switchings: the case of Gaussian switchings and two different trigonometric switching functions with compact support, defined by:
\begin{align}
    \chi_\tc{g}(\tau) &= \frac{e^{- \frac{\tau^2}{2 T^2}}}{\sqrt{2\pi}},\label{chiG}\\
    \chi_\tc{c}(\tau) &= \begin{cases}
        \frac{\pi}{2}\cos\left(\frac{\pi \tau}{T}\right)&, |\tau|<T/2
        \\0 &, |\tau|\geq T/2
    \end{cases}\label{chiC}\\
    \chi_\tc{s}(\tau) &= \begin{cases}
        2\cos^2\left(\frac{\pi \tau}{T}\right)&, |\tau|<T/2
        \\0 &, |\tau|\geq T/2
    \end{cases}\label{chiS}
\end{align}
$T$ here denotes the value of the integral over $\tau$ of each of the functions above, and is to be taken as a control of the time scale of the interactions. We then define the dual switching functions as described in Section \ref{sec:duality}:
\begin{align}
    \Tilde{\chi}_\tc{g}(\tau) &= \left|\int_{-\infty}^\tau \dd t \chi_\tc{g}(t) e^{-\ii \Omega t}\right|\\
    &= \frac{T}{2}e^{- \frac{\Omega^2 T^2}{2}}\left|1 + \text{erf}\left(\frac{\tau/T + \ii \Omega T }{\sqrt{2}}\right)\right|,\nonumber\\
    \Tilde{\chi}_\tc{c}(\tau) &= \left|\int_{-\infty}^\tau \dd t \chi_\tc{c}(t) e^{-\ii \Omega t}\right|,\nonumber\\
    \Tilde{\chi}_\tc{s}(\tau) &= \left|\int_{-\infty}^\tau \dd t \chi_\tc{s}(t) e^{-\ii \Omega t}\right|,\nonumber
\end{align}
where we have omitted the analytical expressions for $\tilde{\chi}_\tc{c}(\tau)$ and $\tilde{\chi}_\tc{s}(\tau)$, as they are cumbersome and do not bring much insight. In Figs. \ref{fig:Gaussian}, \ref{fig:Compact}, and \ref{fig:CompactSmooth} we plot the functions $\Omega \tilde{\chi}_\tc{g}(\tau)$, $\Omega \tilde{\chi}_\tc{c}(\tau)$, and  $\Omega \tilde{\chi}_\tc{s}(\tau)$ for different values of $\Omega$, so that these can be compared with $\chi_\tc{g}(\tau)$, $\chi_\tc{c}(\tau)$, and $\chi_\tc{s}(\tau)$.

In Fig. \ref{fig:Gaussian} we can see that for small $\Omega$ ($\Omega T  = 1$), the function $\tilde{\chi}_{\tc{g}}(\tau)$ is not localized, and varies from $0$ to a constant. This is because for $\Omega T  \ll 1$, $\tilde{\chi}_{\tc{g}}(\tau)$ is simply the integral of a positive function. This means that our duality will not hold for such a small value of $\Omega$.  As $\Omega$ increases, $\tilde{\chi}_\tc{g}(\tau)$ becomes closer to $0$ for larger values of $\tau/T$, and it gets closer and closer to the function $\chi_\tc{g}(\tau)$, as Theorem 1 states. What is remarkable about the case of the Gaussian switching is that for relatively small gaps ($\Omega T \approx 5$) we already obtain a remarkable similarity between the dual switching function $\tilde{\chi}_\tc{g}(\tau)$ and $\chi_\tc{g}(\tau)$. Indeed, we find that for $\Omega T = 5$, the relative $L^1$ distance between the functions is approximately $2.1\%$. This falls to $0.5\%$ for $\Omega T = 10$. Gaussian switching functions are also the most common choice of switching function, and are used\footnote{Notice that in these examples, multiple detectors are used. However, as we will see in Section \ref{sec:harvestingDuality}, the duality also holds for multiple detectors}, for instance, in~\cite{ElModeEstaLouko,Salton:2014jaa,topology,Pozas-Kerstjens:2015,HollowShell,Ng1,Petar,UDWAdS,EricksonZero,freefall,twist2022,diki,hector,hectorChicken}. Specifically, the references~\cite{Pozas-Kerstjens:2015,hectorChicken} consider $\Omega T > 5$ in the parameter space studied with UDW detectors linearly coupled to a quantum field. This implies that their results directly carry on to the case of derivative coupled detectors.

In Fig. \ref{fig:Compact} we plot the function $\tilde{\chi}_\tc{c}(\tau)$ for  different values of $\Omega$. For small values of $\Omega$ we again see that the function starts at $0$ and becomes constant, as it is simply the integral of a positive compactly supported function. However, as $\Omega$ increases, $\tilde{\chi}_\tc{c}(\tau)$ converges to ${\chi}_\tc{c}(\tau)$, becoming localized, and being in the regime where the duality holds. Nonetheless, notice the effect that the non-smoothness of  ${\chi}_\tc{c}(\tau)$ has in this case, which makes the convergence much slower. In this case $\Omega T = 100$ still does not erase the tail of $\tilde{\chi}_\tc{c}(\tau)$ completely, and one requires to go to $\Omega T = 1000$. Overall, for non-smooth switching functions the duality is only expected to hold for $\Omega T \gg 1$, which is usually not very physically relevant.

In Fig. \ref{fig:CompactSmooth} we plot the function $\tilde{\chi}_\tc{s}(\tau)$ for  different values of $\Omega$. Our main goal with this example is to showcase that the smoother the function, the lower value of the energy gap required so that $\Omega\tilde{\chi}(\tau) \approx \chi(\tau)$. For instance, in Fig. \ref{fig:CompactSmooth} we see that the approximation approximately holds for $\Omega T = 50$. Notice that this is much smaller than the values found for $\tilde{\chi}_{\tc{c}}$, which has non-continuous derivatives. 

We have also checked that for large $\Omega$, the phase  functions $\theta(\tau)$ in Eq. \eqref{dualityValues} behave as $\theta(\tau) \approx \Omega \tau - \pi/2$ for all models. This convergence happens much faster than the convergence $\Omega \tilde{\chi}(\tau)\rightarrow \chi(\tau)$.

Overall, the examples of this section are useful to check under which conditions it is possible to extrapolate the results obtained from detectors linearly coupled to the field to the results of detectors coupled with the field's derivative. We saw here that the exact values of detector gaps that allow such duality to approximately hold depend explicitly on the shape of the switching function, and we can infer that smooth switchings will allow the duality to hold for smaller gaps. In general, if a computation with UDW detectors is performed with linearly coupled detectors that satisfy conditions \ref{cond1} and \ref{cond2}, these results will also be valid for the derivative coupling using $\tilde{\lambda} = \lambda $ and the same switching and detector gap if:
\begin{equation}
    \Omega\left|\int_{-\infty}^\tau dt\,e^{-\ii \Omega t}\chi(t)\right| -  \chi(\tau)\ll 1, \ \forall\,\tau.
\end{equation}
That is, if $\Omega \tilde{\chi}(\tau)\approx \chi(\tau)$.


\section{Entanglement Harvesting}\label{sec:harvesting}

In this section, we are going to review the protocol of entanglement harvesting, a notorious example of how UDW-like models have opened the path for the exploration of new phenomena within the framework of QFT in curved spacetimes. Specifically, by coupling two local probes to a quantum field, one can learn how the geometry and topology of spacetime can affect the entanglement structure that is present in the field correlations~\cite{topology,geometry}. For entanglement harvesting in its standard form,  we need to consider two detectors, $A$ and $B$, following timelike trajectories $\mf z_{\tc{a}}(\tau_{\tc{a}})$ and $\mf z_{\tc{b}}(\tau_{\tc{b}})$ in the spacetime ${\mf M}$, which we assume to be parametrized by their respective proper times. The free Hamiltonians of both of the detectors take the same form as before in their proper frames, i.e.
\begin{equation}
    \hat{H}_{\tc{d}, j} = \Omega_{j} \hat{\sigma}_j^{+} \hat{\sigma}_j^{-},
    \label{H_DAB}
\end{equation}
for $j = \tc{A},\tc{B}$.

Then, we couple the detectors with a scalar, real quantum field $\hat{\phi}(\mf x)$. This way, the interaction Hamiltonian density takes the form
\begin{equation}
    \hat{h}_{I}(\mf x) = \lambda (\Lambda_{\tc{a}}(\mf x) \hat{\mu}_{\tc{a}}(\tau_\tc{a}) + \Lambda_{\tc{b}}(\mf x) \hat{\mu}_{\tc{b}}(\tau_\tc{b}))\hat{\phi}(\mf x),
    \label{H_I_harvesting}
\end{equation}
where
\begin{equation}
    \hat{\mu}_{j}(\tau_j) = e^{\ii \Omega_{j} \tau_{j}} \hat{\sigma}^{+}_{j} + e^{-\ii \Omega_{j} \tau_j} \hat{\sigma}^{-}_{j}.
\end{equation}
Moreover, we will assume here that the spacetime smearing functions $\Lambda_\tc{a}(\mf x)$ and $\Lambda_{\tc b}(\mf x)$ have non-overlapping support. Under this assumptions, the FNC times $\tau_\tc{a}(\mf x)$ and $\tau_\tc{b}(\mf x)$ locally defined around each curve can be written unambiguously in Eq. \eqref{H_I_harvesting}. Also notice that the results of~\cite{us2} allow any notion of time ordering to be picked when computing the time evolution operator to second order in $\lambda$.

For the initial state of the full system, we can set both detectors to their ground state, and again assume that the field starts in any Gaussian state $\hat{\rho}_{\phi, 0}$. So, we write the initial state as
\begin{equation}
    \hat{\rho}_{0} = |g_{\tc{a}} \rangle \langle g_\tc{a}| \otimes |g_\tc{b} \rangle \langle g_\tc{b}| \otimes \hat{\rho}_{\phi, 0}.
    \label{rho_0_harvesting}
\end{equation}
Notice that with the choice above, the detectors' initial state \mbox{$\hat{\rho}_{\tc{ab}, 0} = |g_{\tc a} \rangle \langle g_{\tc a}| \otimes |g_{\tc b} \rangle \langle g_{\tc b}|$} is a non-entangled state. However, this might change after they interact with the field. Indeed, using the time evolution operator \eqref{U_I} with the  Hamiltonian density displayed on Eq. \eqref{H_I_harvesting}, we can write the final state of the detectors as $\hat{\rho}_{\tc{ab}} =  \Tr_{\phi}[\hat{U}_{I} \hat{\rho}_{0} \hat{U}_{I}^{\dagger}]$, or, in matrix form,
\begin{equation}
 \hat{\rho}_{\tc{ab}} = \begin{bmatrix}
{ 1 - {\cal L}_{\tc{aa}} - {\cal L}_{\tc{bb}}} & 0 & 0 & {\cal M}^{*} \\
0 & {\cal L}_{\tc{bb}}& {\cal L}_{\tc{ba}} & 0 \\
0 & {\cal L}_{\tc{ab}} & {\cal L}_{\tc{aa}} & 0\\
{\cal M} & 0 & 0 & 0\\ 
\end{bmatrix} 
 +  {\cal O}(\lambda ^{4}),
\label{rho_AB_harvesting}
\end{equation}
where
\begin{align}
    {\cal L}_{ij} &=  \lambda^2 \int{\dd V \dd V'e^{-\ii (\Omega_{i} \tau_i-\Omega_{j}\tau_j')} \Lambda_{i}(\mf x) \Lambda_{j}(\mf x')W(\mf x, \mf x')},\label{Lij_harvesting} \\
    {\cal M} &= -\lambda^2 \int{\dd V \dd V'e^ {\ii( \Omega_{\tc{a}}\tau_\tc{a}+\Omega_{\tc{b}} \tau_\tc{b}')} \Lambda_{\tc{a}}(\mf x)\Lambda_{\tc{b}}(\mf x')G_{F}(\mf x, \mf x')}.
    \label{M_harvesting}
\end{align}
In the last equation, $G_{F}(\mf x, \mf x')$ stands for the {\it Feynman propagator}, which can be written as
\begin{equation}
\begin{aligned}
    G_{F}(\mf x, \mf x') &= \Tr[{\cal T}\hat{\phi}(\mf x)  \hat{\phi}(\mf x') \hat{\rho}_{\phi,0}]  \\&= \theta(t - t') W(\mf x, \mf x') + 
    \theta(t' - t) W(\mf x', \mf x),
\end{aligned}
\end{equation}
where $\theta(t)$ stands for the Heaviside step function, and $t$ denotes any time coordinate.

In order to measure the amount of entanglement between the two detectors, we can use the {\it negativity}, a trustworthy entanglement quantifier for systems of two qubits~\cite{VidalNegativity}. Such a measure has also been a standard choice among works that explore entanglement harvesting (e.g. \cite{Pozas-Kerstjens:2015,Pozas2016,ericksonNew,hectorChicken}) mainly because it is well defined and easy to compute even for higher dimensional quantum systems. For a density operator $\hat{\rho}$ which describes the state of a bipartite system ${\cal H}_{\tc{a}} \otimes {\cal H}_{\tc{b}}$, the negativity ${\cal N}(\hat{\rho})$ of the quantum state $\hat{\rho}$ 
is given by the absolute sum of the {\it negative eigenvalues} of $\hat{\rho}^{t_{\tc{a}}}$, where $\hat{\rho}^{t_{\tc{a}}}$ denotes the partial transpose of $\hat{\rho}$ with respect to ${\cal H}_{\tc{a}}$. In particular, for $\hat{\rho}_{\tc{ab}}$ we obtain the following result for the negativity:

\begin{equation}
   {\cal N}(\hat{\rho}_{\tc{ab}}) =  \max\{0, {\cal V}\},
\end{equation}
where
\begin{equation}
    {\cal V} = \sqrt{|{\cal M}|^{2} +\left(\frac{{\cal L}_{\tc{aa}}
    - {\cal L}_{\tc{bb}}}{2}\right)^2 } - \frac{{\cal L}_{\tc{aa}} + {\cal L}_{\tc{bb}}}{2}.
\end{equation}
Notice that, in general, the negativity will be a competition between the non-local term ${\cal M}$, and the terms that depend only on each detector separately (${\cal L}_{\tc{aa}}$ and ${\cal L}_{\tc{bb}}$). For the special case where $\mathcal{L}_{\tc{aa}} = \mathcal{L}_{\tc{bb}} = \mathcal{L}$ (such as identical inertial detectors in Minkowski spacetime), the negativity reduces to

\begin{equation}
    {\cal N}(\hat{\rho}_{\tc{ab}}) = \max\{0, |{\cal M}| - {\cal L}\}.
\end{equation}

There will be entanglement between the detectors $A$ and $B$ whenever the physical configuration is such that ${\cal V} > 0$. However, it is important to acknowledge that there are two possible sources for the entanglement ~\cite{ericksonNew,quantClass}: communication-mediated entanglement ({\it signalling}) or extraction of entanglement from pre-existing correlations of the field ({\it harvesting}). When the detectors are causally connected, ${\cal N}(\hat{\rho}_{\tc{ab}})$ will have contribution from both sources. Nonetheless, if we consider the detectors to be spacelike separated, then we can be sure that all acquired entanglement is due to true entanglement harvesting, as there cannot be communication between them in this case. We refer the reader to~\cite{ericksonNew,quantClass} for more details.

\section{Duality between linear coupling and derivative coupling in Entanglement Harvesting}\label{sec:harvestingDuality}

    
    

In this section, we are going to discuss how our results about duality between the amplitude and the derivative coupling translate to the case where two detectors interact with the field. This case is relevant for different communication protocols, as well as for entanglement harvesting.

Consider two timelike trajectories $\mf z_\tc{a}(\tau_\tc{a})$ and $\mf z_{\tc{b}}(\tau_\tc{b})$ parametrized by proper time, and consider two particle detectors which interact with a quantum field according to the model described in Section \ref{sec:harvesting}. We wish to compare this case with the case in which the detectors couple according to the derivative coupling.
In the derivative coupling case we write the scalar interaction Hamiltonian density as
\begin{equation}
    \hat{\Tilde{h}}_{\tc{I}}(\mf x) = \hat{\Tilde{h}}_{\tc{I}, \tc{a}}(\mf x) + \hat{\Tilde{h}}_{\tc{I}, \tc{b}}(\mf x),
\end{equation}
with
\begin{equation}
    \hat{\Tilde{h}}_{\tc{I}, j}(\tau) = \lambda\Tilde{\Lambda}_{j}(\mf x)(e^{\ii \theta_j(\tau_j)}\hat{\sigma}^{+}_{j} +  e^{-\ii \theta_j(\tau_j)}\hat{\sigma}^{-}_{j}) \partial_{\tau_j} \hat{\phi}(\mf x),
\end{equation}
where we once again consider time varying gaps $\tilde{\Omega}_j(\tau_j)$ for the detectors, such that $\dv{}{\tau_j}\theta_j(\tau_j) = \tilde{\Omega}_j(\tau_j)$. Notice that in this scenario, each detector couples to the derivative of the field with respect to its own proper time, and that the interaction Hamiltonian densities above are only defined locally around each detector. 

Now, provided that we prepare our system in the ground state ($\hat{\rho}_{\tc{ab},0} = \ket{g_\tc{a}}\!\!\bra{g_\tc{a}}\otimes\ket{g_\tc{b}}\!\!\bra{g_\tc{b}}$) and evolve it similarly to what was done in Section~\ref{sec:harvesting}, the reduced state $\hat{\rho}_{\tc{ab}}$ of the local probes will have the same form as of Eq. \eqref{rho_AB_harvesting} for both choices of coupling (amplitude or derivative). For the case of the amplitude coupling, the terms of the density matrix $\hat{\rho}_{\tc{ab}}$ are given by Eqs. \eqref{Lij_harvesting} and \eqref{M_harvesting}, while in the case of the derivative coupling, the final density matrix is still given by Eq. \eqref{rho_AB_harvesting} with the replacements $\mathcal{L}_{ij} \mapsto \Tilde{\mathcal{L}}_{ij}$ and $\mathcal{M}\mapsto \Tilde{\mathcal{M}}$, which can be written as

\begin{align}
   \Tilde{{\mathcal{L}}}_{ij} &=  \Tilde{\lambda}^2 \int{\dd V \dd V' \Tilde{\Lambda}_{i}(\mf x) \Tilde{\Lambda}_{j}(\mf x')e^{-\ii(\theta_i - \theta_j')}\Tilde{W}_{ij}(\mf x, \mf x')}, \nonumber\\
    \Tilde{{\cal M}} &= -\Tilde{\lambda}^2 \int{\dd V \dd V' \Tilde{\Lambda}_{\tc{a}}(\mf x) \Tilde{\Lambda}_{\tc{b}}(\mf x' )e^{\ii (\theta_\tc{a} + \theta_\tc{b}')}\Tilde{G}_{F}(\mf x, \mf x')}
    \label{M_tilda_AB_original}
\end{align}
and we define
\begin{align}
    \Tilde{W}_{ij}(\mf x, \mf x') &= \langle \partial_{\tau_i} \hat{\phi}(\mf x) \partial_{\tau_j'} \hat{\phi}(\mf x)\rangle = \partial_{\tau_i}\partial_{\tau_j'} W(\mf x, \mf x'),\\
    \Tilde{G}_{F}(\mf x, \mf x') &= \langle\mathcal{T}\partial_{\tau_\tc{a}}\hat{\phi}(\mf x) \partial_{\tau_\tc{b}'}\hat{\phi}(\mf x')\rangle\label{GFder} \\
    &=\theta(t - t') \Tilde{W}_{\tc{ab}}(\mf x, \mf x') + \theta(t' - t) \Tilde{W}_{\tc{ba}}(\mf x', \mf x).\nonumber
\end{align}
The definition above makes it so that the relevant proper time parameter ($\tau_\tc{a}$ or $\tau_\tc{b}$) is used to differentiate the field in the region corresponding to each spacetime smearing function ($\Lambda_{\tc{a}}$ or $\Lambda_{\tc{b}}$). In particular, with this convention we can perform integration by parts in the $ \Tilde{{\mathcal{L}}}_{ij}$ terms by following exactly the same method described in Section \ref{sec:duality}. That is, if conditions \ref{cond1}-\ref{cond3} are satisfied both for detector $A$ and for detector $B$, we have $\mathcal{L}_{ij} = \tilde{\mathcal{L}}_{ij}$.

The next step is to check how ${\cal M}$ and  $\Tilde{{\cal M}}$ relate in this case, and whether it is possible to integrate $\tilde{\mathcal{M}}$ by parts without picking up extra terms. In Appendix \ref{app}, we show that under the assumption of divergencelessness for the vector fields $\partial_{\tau_\tc{a}}$ and $\partial_{\tau_\tc{b}}$, it is possible to perform a similar integration by parts for the $\tilde{\mathcal{M}}$ term, so that it is given by
\begin{equation}
 \Tilde{\cal M} = -\Tilde{\lambda}^2 \int{\dd V \dd V' \partial_{\tau} (\Tilde{\Lambda}_{\tc{a}}(\mf x)e^{\ii \theta_{\tc{a}}}) \partial_{\tau'}(\Tilde{\Lambda}_{\tc{b}}(\mf x')}e^{\ii \theta_\tc{b}'}) G_{F}(\mf x, \mf x').
 \label{M_tilda_AB_final}
\end{equation}
It might in principle look like the Heaviside thetas in Eq.~\eqref{GFder} might add extra terms when integrating the term $\Tilde{\mathcal{M}}$ by parts. However, these terms end up evaluating to equal time commutator of field observables, which vanish under the assumption of microcausality for the field. The details can be found in Appendix \ref{app}.

Overall, we conclude that the duality we stated in Section \ref{sec:duality} also holds for multiple detectors coupled to the field. That is, for sufficiently large gaps, derivative coupled detectors with switching functions \mbox{$\tilde{\chi}_j(\tau) = \chi_j(\tau)/\Omega_j$} will give the exact same result as an amplitude coupled field would. This implies that any study considered in the literature of quantum communication protocols using multiple localized particle detectors coupled to the field amplitude directly generalize to derivative coupled detectors, provided that \ref{cond1}-\ref{cond3} are satisfied for both detectors.

\section{Conclusions}\label{sec:conclusions}

We have studied particle detectors linearly coupled to the amplitude of a quantum field $\hat{\phi}(\mf x)$ and to its derivative $\partial_\tau\hat{\phi}(\mf x)$ in a curved background spacetime.
We established a duality between the amplitude coupling and the derivative coupling under the assumptions of \ref{cond1}) rigidity, \ref{cond2}) constant local spacetime volume around the trajectories, \ref{cond3}) limit of large energy gaps. We have shown that in this case, a mere rescaling of the coupling constant (or equivalently, of the spacetime smearing function) makes it so that both models yield the exact same results.
 

We have also shown that the duality holds in the case where two detectors are coupled to the quantum field. This makes it so that any conclusion that can be made for detectors with sufficiently large energy gaps coupled to the field amplitude are also valid for detectors coupled to the field's momentum, with a mere rescaling of the results by the energy gap. This result points towards an universality in particle detector models, allowing for the conjecture that in the limit of large gaps, particle detectors will share the same behaviour, regardless of which field operator they are coupled to.

We also studied explicit examples in order to determine how large the detector's gap has to be in order for the duality to approximately hold. We found that the smoother the switching function, the smaller the detector gap has to be so that the derivative coupling shares the same behaviour as the amplitude coupling model.

Overall, our results represent a universal behaviour of particle detectors in the regime of large gaps, where the final state of the detector is independent on whether it is coupled to the field, or to its conjugate momentum. The duality presented here can then be used to extrapolate future and past results of entanglement harvesting 
using amplitude coupled detectors to the case where the detectors are instead coupled to the field's momentum. 

\begin{acknowledgements}
The authors thank Prof. Jorma Louko for insightful discussions about IR regulators and assymptotic exansions. TRP acknowledges support from the Natural Sciences and Engineering Research Council of Canada (NSERC) via the Vanier Canada Graduate Scholarship. MHZ also thanks Prof. Achim Kempf and Prof. Eduardo Mart\'in-Mart\'inez’s funding through their NSERC Discovery grants. Research at Perimeter Institute is supported in part by the Government of Canada through the Department of Innovation, Science and Industry Canada and by the Province of Ontario through the Ministry of Colleges and Universities. Perimeter Institute and the University of Waterloo are situated on the Haldimand Tract, land that was promised to the Haudenosaunee of the Six Nations of the Grand River, and is within the territory of the Neutral, Anishinaabe, and Haudenosaunee people.
\end{acknowledgements}

\onecolumngrid

\appendix

\section{Integration of the Non-local term $\mathcal{M}$}\label{app}

In this Appendix, we show in detail how Eq. \eqref{M_tilda_AB_final} can be obtained from Eq. \eqref{M_tilda_AB_original}. In the calculations below, we employ the following definition

\begin{equation}
    f_{j}(\mf x) = \frac{\partial t(\mf x)}{\partial \tau_{j}} , \ j=A,B.
\end{equation}

First of all, from the definition of the Feynman propagator the $\Tilde{\mathcal M}$ term in Eq. \eqref{M_tilda_AB_original} can be written as:

\begin{equation}
    \Tilde{\mathcal{M}} = - \lambda^2 \int \dd V \dd V' \Tilde{\Lambda}_\tc{a}(\mf x)\Tilde{\Lambda}_\tc{b}(\mf x')(\theta(t - t')\partial_{\tau_{\tc{a}}} \partial_{\tau'_{\tc{b}}}W(\mf x, \mf x')+\theta(t' - t)\partial_{\tau_{\tc{a}}} \partial_{\tau'_{\tc{b}}}W(\mf x', \mf x)).
\end{equation} 
    
Then, we perform integration by parts on the variable $\tau_{\tc{a}}$. Since $\partial_{\tau_{\tc{a}}}\theta(t - t') = f_{\tc{a}}(\mf x) \delta(t -t')$, we have

\begin{align}
       \Tilde{\mathcal{M}} = \lambda^2 \int \dd V \dd V' &\partial_{\tau_{\tc{a}}}\Tilde{\Lambda}_\tc{a}(\mf x)\Tilde{\Lambda}_\tc{b}(\mf x')\theta(t - t') \partial_{\tau'_{\tc{b}}}W(\mf x, \mf x')+\Tilde{\Lambda}_\tc{a}(\mf x)\Tilde{\Lambda}_\tc{b}(\mf x')f_{\tc{a}}(\mf x)\delta(t - t') \partial_{\tau'_{\tc{b}}}W(\mf x, \mf x')\nonumber\\
     &+\partial_{\tau_{\tc{a}}}\Tilde{\Lambda}_\tc{a}(\mf x)\Tilde{\Lambda}_\tc{b}(\mf x')\theta(t' - t) \partial_{\tau'_{\tc{b}}}W(\mf x', \mf x)-\Tilde{\Lambda}_\tc{a}(\mf x)\Tilde{\Lambda}_\tc{b}(\mf x')f_{\tc{a}}(\mf x)\delta(t' - t) \partial_{\tau'_{\tc{b}}}W(\mf x', \mf x)\nonumber.
\end{align}

Following the same procedure with the variable $\tau'_{\tc{b}}$, it follows that

\begin{align}
    \Tilde{\mathcal{M}}  
     =  -\lambda^2 \int \dd V \dd V' &\partial_{\tau_{\tc{a}}}\Tilde{\Lambda}_\tc{a}(\mf x)\partial_{\tau'_{\tc{b}}}\Tilde{\Lambda}_\tc{b}(\mf x')\theta(t - t') W(\mf x, \mf x')-\partial_{\tau_{\tc{a}}}\Tilde{\Lambda}_\tc{a}(\mf x)\Tilde{\Lambda}_\tc{b}(\mf x')f_{\tc{b}}(\mf x')\delta(t - t') W(\mf x, \mf x')\nonumber\\
     &+\Tilde{\Lambda}_\tc{a}(\mf x)\partial_{\tau'_{\tc{b}}}\Tilde{\Lambda}_\tc{b}(\mf x')f_{\tc{a}}(\mf x)\delta(t - t') W(\mf x, \mf x')+\Tilde{\Lambda}_\tc{a}(\mf x)\Tilde{\Lambda}_\tc{b}(\mf x')f_{\tc{a}}(\mf x)f_{\tc{b}}(\mf x')\partial_{t'}\delta(t - t') W(\mf x, \mf x')\nonumber\\
     &+\partial_{\tau_{\tc{a}}}\Tilde{\Lambda}_\tc{a}(\mf x)\partial_{\tau'_{\tc{b}}}\Tilde{\Lambda}_\tc{b}(\mf x')\theta(t' - t) W(\mf x', \mf x)+\partial_{\tau_{\tc{a}}}\Tilde{\Lambda}_\tc{a}(\mf x)\Tilde{\Lambda}_\tc{b}(\mf x')f_{\tc{b}}(\mf x')\delta(t' - t) W(\mf x', \mf x)\nonumber\\
     &-\Tilde{\Lambda}_\tc{a}(\mf x)\partial_{\tau'_{\tc{b}}}\Tilde{\Lambda}_\tc{b}(\mf x')f_{\tc{a}}(\mf x)\delta(t' - t) W(\mf x', \mf x)-\Tilde{\Lambda}_\tc{a}(\mf x)\Tilde{\Lambda}_\tc{b}(\mf x')f_{\tc{a}}(\mf x)f_{\tc{b}}(\mf x')\partial_{t'}\delta(t' - t)W(\mf x', \mf x).
\end{align}

Next, we organize the terms and simplify the full expression by using  $\delta(t - t') = \delta(t'  - t)$, $\partial_{t'}\delta(t - t') = -\delta'(t - t')$, and $\partial_{t'}\delta(t' - t) = \delta'(t' - t) = -\delta'(t - t')$. This way, the expression can be cast into

\begin{align}
    \Tilde{\mathcal{M}} 
     =  -\lambda^2 \int \dd V \dd V' &\partial_{\tau_{\tc{a}}}\Tilde{\Lambda}_\tc{a}(\mf x)\partial_{\tau'_{\tc{b}}}\Tilde{\Lambda}_\tc{b}(\mf x')G_F(\mf x, \mf x')\nonumber\\
     &+\delta(t - t')(f_{\tc{a}}(\mf x)\Tilde{\Lambda}_\tc{a}(\mf x)\partial_{\tau'_{\tc{b}}}\Tilde{\Lambda}_\tc{b}(\mf x') - f_{\tc{b}}(\mf x')\partial_{\tau_{\tc{a}}}\Tilde{\Lambda}_\tc{a}(\mf x)\Tilde{\Lambda}_\tc{b}(\mf x')) (W(\mf x, \mf x')- W(\mf x', \mf x))\nonumber\\
     &+\delta'(t - t')f_{\tc{a}}(\mf x)f_{\tc{b}}(\mf x')\Tilde{\Lambda}_\tc{a}(\mf x)\Tilde{\Lambda}_\tc{b}(\mf x')(W(\mf x', \mf x)- W(\mf x, \mf x'))\nonumber\\
\end{align}

Finally, notice that the remaining Wightman function terms are equivalent to the average value of the commutator $[\hat{\phi}(\mf x), \hat{\phi}(\mf x')]$ evaluated over the initial state of the field, $\hat{\rho}_{\phi, 0}$. Writing this average simply as $\langle [\hat{\phi}(\mf x), \hat{\phi}(\mf x')]\rangle$, we have

\begin{align}
   \Tilde{\mathcal{M}} 
     =  -\lambda^2 \int \dd V \dd V' &\partial_{\tau_{\tc{a}}}\Tilde{\Lambda}_\tc{a}(\mf x)\partial_{\tau'_{\tc{b}}}\Tilde{\Lambda}_\tc{b}(\mf x')G_F(\mf x, \mf x')\nonumber\\
     &+ f_{\tc{a}}(\mf x)f_{\tc{b}}(\mf x')\Tilde{\Lambda}_\tc{a}(\mf x)\Tilde{\Lambda}_\tc{b}(\mf x')\left(\delta(t - t')\left(\frac{\partial_{\tau'_{\tc{b}}}\Tilde{\Lambda}_\tc{b}(\mf x')}{f_{\tc{b}}(\mf x')\Tilde{\Lambda}_\tc{b}(\mf x')} - \frac{\partial_{\tau_{\tc{a}}}\Tilde{\Lambda}_\tc{a}(\mf x)}{f_{\tc{a}}(\mf x)\Tilde{\Lambda}_\tc{a}(\mf x)}\right)-\delta'(t - t')\right) \langle [\hat{\phi}(\mf x), \hat{\phi}(\mf x')]\rangle.\nonumber
\end{align}
Notice that the terms which are integrals of the commutator all have deltas or derivatives of deltas evaluating them. This means that each of these terms will evaluate to integrals of the equal time commutator of observables, which are zero under the assumption of microcausality (observables evaluated at spacelike separated points vanish). Therefore we find,

\begin{equation}
    \Tilde{\mathcal{M}} 
     =  -\lambda^2 \int \dd V \dd V' \partial_{\tau_{\tc{a}}}\Tilde{\Lambda}_\tc{a}(\mf x)\partial_{\tau'_{\tc{b}}}\Tilde{\Lambda}_\tc{b}(\mf x')G_F(\mf x, \mf x'),
\end{equation} as stated in Eq. \eqref{M_tilda_AB_final}.

\twocolumngrid

\bibliography{references}

\end{document}